\documentclass[journal]{IEEEtran}

\usepackage[percent]{overpic}
\usepackage{amsthm} 
\usepackage{amsmath,amssymb}
\allowdisplaybreaks
\AtBeginDocument{\setlength{\abovedisplayskip}{4pt plus 2pt minus 1pt}\setlength{\belowdisplayskip}{4pt plus 2pt minus 1pt}\setlength{\abovedisplayshortskip}{2.5pt plus 1.5pt minus 1pt}\setlength{\belowdisplayshortskip}{2.5pt plus 1.5pt minus 1pt}}
\usepackage{booktabs}
\usepackage{siunitx}
\usepackage{braket}

\theoremstyle{definition}

\theoremstyle{remark}

\usepackage{lineno}
\usepackage[caption=false,font=footnotesize]{subfig}
\newenvironment{subfigure}[2][]{\begin{minipage}{#2}}{\end{minipage}} 
\usepackage{graphicx}
\usepackage{url}
\usepackage{float}

\begin{document}
\title{
Phase-Stabilization-Free Information Processing on Unstable Photonic Processors}

\author{Gökhan Elmas, Igor A Litvin and Janis Nötzel\thanks{Emmy-Noether Group Theoretical Quantum Systems Design, Technical University of Munich, Munich, Germany}\thanks{gokhan.elmas@tum.de}}

\maketitle
\begin{abstract}
Hadamard-coded optical receivers commonly require stable interference conditions for absolute codeword discrimination. Here we encode four symbols as differential Hadamard operations: no phase change or a $\pi$ shift on exactly two of three non-reference input modes. The information is carried by the selected phase-change operation, rather than the absolute optical codeword. We investigate a fixed eight-port intensity-only receiver that reconstructs six coupled trigonometric observables from simultaneous power measurements. Their deterministic sign transformations allow an analytical decoder to identify the applied operation from successive measurements without continuous optical phase stabilization or knowledge of the initial absolute codeword label. An earlier measured signature and sufficiently limited inter-measurement drift remain necessary. We establish exact observable reconstruction, characterize exceptional phase configurations and conditional noise sensitivity, and compare ideal four- and eight-port detection under an equal-photon budget. Experiments on programmable photonic processors yield 96.19\% analytical decoding accuracy; a separate machine-learning analysis reports 98.31\% on a held-out experimental trace. The results demonstrate differential, intensity-only Hadamard operation decoding under the tested conditions.
\end{abstract}


\section{Introduction}

Hadamard encoding provides orthogonal codewords for optical communication and signal processing, enabling low-complexity encoding and decoding with useful discrimination properties~\cite{Li,Hadamard_CDMA,Noshad}. The four-dimensional Hadamard matrix ($H_4$) provides a compact realization of these properties using a relatively small number of optical modes. Advances in integrated and quantum photonics have made complex linear-optical transformations increasingly accessible on programmable chips~\cite{Luo,Flamini2019,aghaee,maring}. In an ideal phase-stable receiver, the inverse Hadamard transformation concentrates each codeword in a corresponding output port, permitting direct intensity-based discrimination. This ideal behavior becomes less reliable when the encoded modes acquire different phase shifts.

In practice, phase fluctuations arising from environmental perturbations, fabrication tolerances, and propagation effects~\cite{Elmas2025} disturb the relative phases required for coherent interference and can make codeword discrimination unreliable. Active phase stabilization and tracking can mitigate these effects~\cite{Litvin2025,Litvin2}. Recent demonstrations include on-chip feedback for four-input Hadamard transformations~\cite{LitvinMultiInput2026} and Hadamard-based joint detection using separate programmable processors for codeword generation and decoding~\cite{LitvinJointDetection2026}. Although effective, such approaches introduce feedback, calibration, and control requirements. These requirements become particularly relevant for multi-input systems and changing channel conditions. A decoder that remains optically fixed during operation could reduce feedback and recalibration overhead, even though maintaining a thermo-optically programmed transformation may still consume electrical power.

Phase instability is also a challenge in quantum photonic platforms employing squeezed optical states. For example, the Borealis processor used actively phase-stabilized interferometric loops for large-scale Gaussian boson sampling~\cite{Madsen2022}, while recent theoretical studies have investigated the effects of input dephasing on such systems~\cite{Paryzkova2026}. For optical decoding, obtaining complementary phase information simultaneously is particularly useful in view of the limitations of sequential quadrature measurements~\cite{ElmasQuadrature2026}. The optical Green Machine uses a passive Hadamard transformation to turn suitable phase-coded inputs into intensity-distinguishable outputs~\cite{Cui}, while universal interferometers allow more general programmable transformations~\cite{Reck1994,Qiang2018}. Related developments in photonic information processing and integrated devices provide further context~\cite{Harris,chip1}. Nevertheless, conventional four-port output-power concentration can become ambiguous at certain unknown relative-phase configurations. This motivates a receiver that retains direct power detection but exposes additional, complementary phase signatures without optical reconfiguration between measurements.

Here we use a differential $H_4$ signaling alphabet in which the transmitter encodes each symbol by choosing one of four operations: leaving the relative phases unchanged or adding $\pi$ to exactly two of the three non-reference phases. Thus the encoded information specifies \emph{which phase shifts occur}, not which absolute Hadamard state is occupied. To recover this operation under unknown input-relative phases, we design a fixed eight-port transformation whose simultaneous power measurements allow exact reconstruction of six coupled sine- and cosine-type observables. Their sign changes follow deterministic rules for the four operations, enabling differential identification without continuous optical phase-error feedback or coherent detection. The analytical decoder needs a preceding measured signature and sufficiently limited drift between observations, but not an initial absolute codeword label. We establish the reconstruction map, characterize exceptional phase configurations and noise sensitivity, and present an idealized equal-photon-budget four- versus eight-port comparison. Experiments demonstrate analytical transition decoding, complemented by a separately evaluated machine-learning approach. The contribution is the combined differential signaling rule, eight-port optical construction, and operation identification from intensity-only signatures.

\section{Decoding Under Phase Noise}
\label{sec:decoding-phase-noise}

\subsection{Four-Port Hadamard Decoding}
\label{subsec:H4-summary}
\label{sec:idealH4}

Conventional four-port Hadamard decoding uses the normalized Sylvester-ordered matrix
\begin{equation*}
H_4=\frac12\begin{pmatrix}
1&1&1&1\\
1&-1&1&-1\\
1&1&-1&-1\\
1&-1&-1&1
\end{pmatrix},\qquad H_4^T H_4=H_4^2=I_4.
\end{equation*}
The four conventional Hadamard codewords are equal-amplitude orthogonal states: for a basis vector $\mathbf e_i$, the transmitter may prepare $\mathbf x_i=H_4\mathbf e_i$, and an ideal inverse receiver gives $H_4\mathbf x_i=\mathbf e_i$. Our signaling rule differs from labeling these absolute states. At symbol interval $k$, the transmitter instead selects an \emph{operation} $g_k$ from the four-element alphabet
\begin{equation}
\begin{aligned}
\mathcal G&=\{I,T_{BC},T_{BD},T_{CD}\},\\
\mathbf v_g^T&\in\{(0,0,0),(1,1,0),(1,0,1),(0,1,1)\},
\end{aligned}
\label{eq:differential_alphabet}
\end{equation}
where the entries of $\mathbf v_g$ correspond to the non-reference phases $B,C,D$, respectively. The selected symbol commands $\pi\mathbf v_g$: either no change or a $\pi$ shift on exactly two phases. For the receiver-relative phase triple $\boldsymbol\theta_k=(B_k,C_k,D_k)^T$, successive observations obey
$\boldsymbol\theta_{k+1}=\boldsymbol\theta_k+\pi\mathbf v_{g_k}+\boldsymbol\xi_k\pmod{2\pi}$,
where $\boldsymbol\xi_k$ collects inter-interval phase drift. Each allowed operation maps a Hadamard codeword into another codeword in the absence of drift, but the transmitted information is $g_k$, not the absolute codeword index.

When the four encoded modes acquire relative phase fluctuations, choosing mode~1 as a reference, Bob's input state can instead be written as $\tfrac12(1,e^{iB},e^{iC},e^{iD})^T$, with $B,C,D\in[0,2\pi)$. Here $B$, $C$, and $D$ denote the total relative phases, incorporating both the encoded $0$ or $\pi$ offsets and any subsequently acquired phase noise.

Applying $H_4$ yields four output powers depending on these relative phases. For example,
\begin{equation}
\begin{aligned}
P_1 = \tfrac{1}{8}\Big(&2+\cos B+\cos C+\cos D\\
&+\cos(B-C)+\cos(B-D)+\cos(C-D)\Big),
\end{aligned}
\end{equation}
while the remaining powers $P_j$ ($j=2,3,4$) follow by analogous sign permutations.

The three nonidentity operations in Eq.~\eqref{eq:differential_alphabet} flip the signs of exactly two Hadamard components, while $I$ leaves all phases unchanged. The $\pi$ shifts change interference terms according to $\cos(\theta+\pi)=-\cos\theta$ and induce known permutations of the conventional four-port powers. The receiver's task is to identify which of these four operations was selected.

\subsection{Codeword Transitions Under Phase Noise}
\label{subsec:H4-noise}

For example, flipping modes 2 and 3 corresponds to
$D_{BC}=\operatorname{diag}(1,-1,-1,1)$. The Hadamard algebra gives
$H_4D_{BC}=\Pi_{BC}H_4$, where $\Pi_{BC}$ reverses the four output
ports. Therefore the output powers change exactly as
$(P_1,P_2,P_3,P_4)\mapsto(P_4,P_3,P_2,P_1)$, regardless of the
initial phases $B,C,D$, provided no additional phase drift occurs.
With an inter-measurement phase increment
$\boldsymbol\Delta=(\Delta B,\Delta C,\Delta D)^T$,
this correspondence is instead accurate up to
$O(\|\boldsymbol\Delta\|_2)$.
The exact permutation matrix and expanded drift expressions are given
in the Supplementary Material. Thus the main limitation is not the
permutation identity itself, but ambiguity of some measured power
patterns and drift relative to the reference observation.

\subsection{Extension to an Eight-Dimensional Framework}
\label{subsec:H8-summary}

The real-valued four-port Hadamard transformation produces cosine-type interference terms but does not directly provide complementary sine-type observables. We therefore embed the four-dimensional Hadamard code into an eight-dimensional passive interferometric architecture that also provides phase-sensitive sine-type information.

The logical Hadamard codeword occupies modes~1--4, corresponding to the four equal-amplitude
components of a single row of $H_4$. The remaining four modes act as ancillary channels and are
initially unpopulated. The input state is thus
\begin{equation}
\begin{aligned}
\mathbf{x}_\text{in} &= \tfrac{1}{2}
\begin{bmatrix}
1 & e^{iB} & e^{iC} & e^{iD} & 0 & 0 & 0 & 0
\end{bmatrix}^{\!T},\\
& B,C,D \in [0,2\pi),
\end{aligned}
\label{eq:h8_input}
\end{equation}
Here $B$, $C$, and $D$ are the effective phases of the populated modes relative to mode~1, including phase variations accumulated after encoding. Throughout the analysis, the optical decoding transformation $U$ is assumed to remain fixed; internal drift of the programmed interferometer is not included in this model.

The decoding operation is implemented by a fixed unitary transformation
$U \in \mathbb{C}^{8\times 8}$,
\begin{equation}
U=\tfrac{1}{2\sqrt{2}}\;
\scalebox{0.7}{$
\begin{bmatrix}
1&-1&-1&i&i&1&-i&-i\\
-1&1&-1&i&-i&1&i&-i\\
i&-i&-i&1&1&i&-1&-1\\
i&-i&i&-1&1&-i&-1&1\\
-i&-i&-i&-1&1&-i&1&-1\\
-i&-i&i&1&1&i&1&1\\
1&1&-1&-i&-i&-1&-i&-i\\
1&1&1&i&-i&1&-i&i
\end{bmatrix}
$}.
\label{eq:h8_unitary}
\end{equation}

Applying $U$ to $\mathbf{x}_\text{in}$ yields eight output amplitudes, whose corresponding
intensities $P_1,\ldots,P_8$ contain both cosine- and sine-type dependencies on the relative
phases. For example,
\begin{equation}\label{eq:pow1}
\begin{aligned} P_1=\tfrac{1}{16}\big(&2-\cos B-\cos C-\sin D\\
&+\cos(B-C)-\sin(B-D)-\sin(C-D)\big),\end{aligned}
\end{equation}
while the remaining output powers follow by analogous combinations.

This transformation makes six trigonometric observables analytically recoverable from the eight intensity measurements. For instance,
\begin{equation}\label{eq:trig_expressions}
\cos B = 4 (P_5 + P_6 + P_7 + P_8) - 2,
\end{equation}
with the other five reconstruction expressions and the full proof given in the Supplementary Material. These six quantities form a linearly recoverable, but phase-constrained, signature.

Complementary trigonometric dependence can help avoid some zero-crossing ambiguities of a cosine-only power pattern, although no phase-uniform margin is guaranteed. The optical decoding transformation is fixed during operation and requires only power detection, not coherent field measurement or active input-phase feedback. In a thermo-optically programmed mesh, holding the fixed transformation may still require electrical bias power.

\subsection{Controlled comparison with conventional four-port decoding}
\label{subsec:controlled-comparison}

The conventional Hadamard receiver and the eight-port decoder both conserve total input power, but their intensity signatures need not have the same distinguishability when the encoded modes acquire unknown relative phases. This can be illustrated without assigning a phase-noise model. In the port convention of the displayed Sylvester matrix in Section~\ref{sec:idealH4}, take
\(B=0,\ C=\pi/2,\ D=-\pi/2\).
The conventional receiver gives the uniform power vector \(\mathbf p_{4}=\tfrac14(1,1,1,1)^T\).
All four Hadamard transitions permute this tuple without changing it, so this particular phase configuration is exactly ambiguous to the conventional intensity-only decoder. Applying the eight-port unitary of Eq.~\eqref{eq:h8_unitary} to the state in Eq.~\eqref{eq:h8_input} instead yields
\(\mathbf p_{8}=\tfrac1{16}(1,1,1,1,1,5,1,5)^T\).
The four candidate eight-port power signatures are distinct, with minimum observable separation $d_{\min}=2\sqrt{2}$ (see the Supplementary Material for the proof). This establishes an explicit advantage at the chosen point, not uniform dominance over every phase and measurement-noise condition.

\begin{figure}[!t]
\centering
\includegraphics[width=0.9\linewidth]{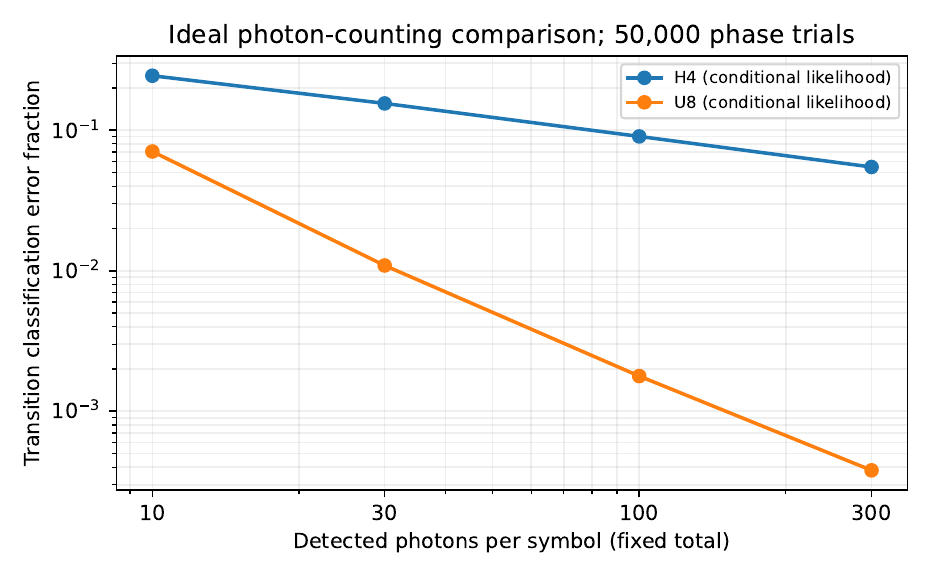}
\caption{Controlled ideal four-port versus eight-port comparison with an unchanged known phase reference, uniform random phase triples, 50,000 trials per count level, equal total detected photons, and conditional multinomial maximum-likelihood decisions. These are simulations, not measured decoding accuracies.}
\label{fig:receiver_comparison}
\end{figure}

To provide a reproducible, equal-photon-budget comparison, we also simulated ideal photon-counting discrimination with the true, unchanged pre-transition phase triple supplied to the decision rule as an ideal reference (an oracle assumption). Each of 50,000 trials draws $B,C,D$ independently and uniformly on $[0,2\pi)$, selects one of four transitions uniformly, and generates $n$ detected photons from the corresponding four- or eight-output multinomial power distribution. For both receivers the same total count $n$ is used, and the transition is assigned to the candidate maximizing the conditional multinomial likelihood. Figure~\ref{fig:receiver_comparison} shows the resulting ideal-model error fractions. At $n=30$, the simulated rates are $0.15510$ for conventional four-port detection and $0.01090$ for the eight-port transformation. The comparison tests the intensity signatures under a specific photon-counting model; it does not reproduce the experimental sequential sign algorithm, incorporate reference-estimation errors or phase changes during an observation window, or establish a hardware-independent advantage. The simulation details and reproducibility code accompany the source files. This reference-conditioned receiver comparison is separate from the reported machine-learning decoder.

\subsection{Sign-Flip Structure and Noise-Robust Decoding}
\label{subsec:sign-structure}

The availability of both cosine- and sine-type observables in the eight-dimensional decoder
enables a discrete, noise-robust representation of Hadamard codeword transitions. As shown
above, the output intensities provide six complementary trigonometric observables, allowing transition signs to be compared through several interference terms. We exploit this structure by focusing on the \emph{signs} of a
small set of analytically reconstructed trigonometric observables.

Specifically, from the measured output powers we reconstruct the six quantities
\(\cos B,\ \cos C,\ \sin D,\ \cos(B-C),\ \sin(B-D),\ \sin(C-D)\)
and define the corresponding sign vector
\begin{equation}
\begin{aligned}
\mathbf{s}(t)=\big(&\mathrm{sgn}[\cos B],\,\mathrm{sgn}[\cos C],\\
&\mathrm{sgn}[\sin D],\,\mathrm{sgn}[\cos(B-C)],\\
&\mathrm{sgn}[\sin(B-D)],\,\mathrm{sgn}[\sin(C-D)]\big)^{T}.
\end{aligned}
\end{equation}

For the operation alphabet in Eq.~\eqref{eq:differential_alphabet}, the observable vector obeys $\mathbf q_{k+1}=S_{g_k}\mathbf q_k$ when the phases do not drift, with the four diagonal sign operators $S_g$ given in the Supplementary Material. The unknown initial phases affect $\mathbf q_k$, but not the operation-dependent sign pattern. For example, choosing $g_k=T_{BC}$ induces

\[
\begin{aligned}
&\cos B \mapsto -\cos B,\quad \cos C \mapsto -\cos C,\\
&\sin(B-D) \mapsto -\sin(B-D),\\
&\sin(C-D) \mapsto -\sin(C-D),
\end{aligned}
\]
while $\sin D$ and $\cos(B-C)$ remain unchanged. The identity operation leaves all six signs unchanged; each of the other operations has its own fixed sign-inversion pattern. With additional drift, $\mathbf q_{k+1}$ generally differs from $S_{g_k}\mathbf q_k$, so the previous measured signature is a temporal reference rather than a known absolute codeword.

This representation can offer a robustness advantage. Except near zero crossings,
the sign of a trigonometric observable is stable under moderate phase noise, even when its
magnitude fluctuates. Moreover, other reconstructed observables may remain sufficiently far from zero to support a transition decision. The implemented analytical strategy therefore uses pattern matching on $\mathbf{s}(t)$ rather than full phase estimation; the amount of noise tolerance depends on the observable magnitudes, reference stability, and detector statistics.

In this framework, the decoded symbol is the estimated transmitter operation $\widehat g_k\in\mathcal G$. It is determined from changes in successive intensity-derived signatures, not from an absolute optical state or explicit phase recovery.

\subsection{Reconstruction and Identifiability}

Let $\mathbf p=(P_1,\ldots,P_8)^T$ and let $\mathbf q$ contain the six
trigonometric observables defined above. The fixed transformation of
Eq.~\eqref{eq:h8_unitary} gives
$\mathbf p=\tfrac18\mathbf 1_8+\tfrac1{16}A\mathbf q$,
where $A$ is an $8\times6$ sign matrix satisfying $A^TA=8I_6$ and
$A^T\mathbf 1_8=0$. Thus the exact observable reconstruction is
$\mathbf q=2A^T\mathbf p$. The complete matrix and explicit
reconstruction formulas appear in the Supplementary Material.

The four candidate signatures of the allowed Hadamard transitions are
pairwise distinct except at 24 isolated input-phase configurations.
This establishes almost-everywhere identifiability, but not a uniform
noise margin. For a known unchanged reference and independent isotropic
Gaussian noise of variance $\sigma_{\mathrm{obs}}^2$ on the reconstructed observables,
a minimum-distance decoder obeys
$P_{\mathrm{error}}\leq\min\{1,\tfrac32\exp[-d_{\min}^2/(8\sigma_{\mathrm{obs}}^2)]\}$.
The precise assumptions, proofs, and correlated-noise extension are
provided in the Supplementary Material; this bound does not directly
apply to the implemented sequential sign decoder under drift.

In the following sections, we validate this decoding strategy experimentally and extend it using data-driven classifiers to further enhance robustness.
\section{Experimental Proof-of-Concept: Hadamard-2 Decoder}

To validate the feasibility of the proposed decoding scheme on real photonic hardware, we implemented a proof-of-concept experiment using a simplified Hadamard-2 decoder. This experiment serves as an initial step toward scalable implementations of higher-dimensional Hadamard codes.

We begin with the normalized $2\times2$ Hadamard matrix
\begin{equation}
H_2=\frac{1}{\sqrt{2}}
\begin{pmatrix}
1 & 1\\
1 & -1
\end{pmatrix}.
\end{equation}
A codeword transition corresponds to a relative
$\pi$-phase shift between the two input arms.

To emulate the principles of the higher-order \(H_4\) decoding scheme in hardware, we extended the transformation to four output modes by embedding the two input channels into a \(4 \times 4\) unitary. The transformation was chosen such that both the cosine and sine of the relative phase appear directly in the output powers:
\begin{equation}
U = \tfrac{1}{2}\;
\scalebox{0.8}{$
\begin{bmatrix}
1 & 1 & 1 & 1 \\
1 & -1 & 1 & -1 \\
1 & -i & -1 & i \\
1 & i & -1 & -i
\end{bmatrix}
$}.
\end{equation}
The columns of \(U\) are orthonormal, ensuring unitarity and conservation of total optical power.

The input state, corresponding to two modes with a relative phase \(\alpha\), is
\(\mathbf{x} = \tfrac{1}{\sqrt{2}}\begin{bmatrix}1 & e^{i\alpha} & 0 & 0\end{bmatrix}^{T}\)
and the output is given by \(\mathbf{y} = U \mathbf{x}\). The measured powers at the four detectors are then
\begin{subequations} \label{eq:Palpha}
\begin{align}
P_{1,2} &= \tfrac{1}{4}\bigl(1 \pm \cos \alpha \bigr), \label{eq:Palpha_ab} \\
P_{3,4} &= \tfrac{1}{4}\bigl(1 \pm \sin \alpha \bigr), \label{eq:Palpha_cd}
\end{align}
\end{subequations}

From these outputs, the relative phase \(\alpha\) can be directly reconstructed:
\begin{subequations} \label{eq:trig_recovery}
\begin{align}
\cos(\alpha) &= 2 \cdot (P_1 - P_2), \label{eq:trig_recovery_a} \\
\sin(\alpha) &= 2 \cdot (P_3 - P_4). \label{eq:trig_recovery_b}
\end{align}
\end{subequations}
This analytic extraction of phase allows the decoder to continuously track \(\alpha\) and identify abrupt \(\pi\)-shifts corresponding to codeword transitions. As in the higher-dimensional case, the decoder selects the trigonometric component (either \(\cos(\alpha)\) or \(\sin(\alpha)\)) with the larger magnitude to mitigate noise sensitivity near zero-crossings.

\subsection{Quadrature Sensitivity to Phase Transitions}

The reconstructed sine and cosine are complementary: a $\pi$-phase
transition changes $(\cos\alpha,\sin\alpha)$ to its negative, so the
magnitude of the jump in this two-observable plane is exactly $2$ for
every initial phase $\alpha$. Moreover,
$\max(|\cos\alpha|,|\sin\alpha|)\geq 1/\sqrt{2}$, ensuring
that at least one observable has a transition magnitude of at least
$\sqrt{2}$. Among pairs of equally normalized sinusoidal observables,
quadrature separation maximizes the worst-case transition response; the
full minimax derivation and its scope are provided in the Supplementary
Material. This does not establish optimality among all possible passive
multi-output receivers.

\begin{figure}[!t]
    \centering
    \includegraphics[
        width=\columnwidth,
        height=0.6\textheight,
        keepaspectratio
    ]{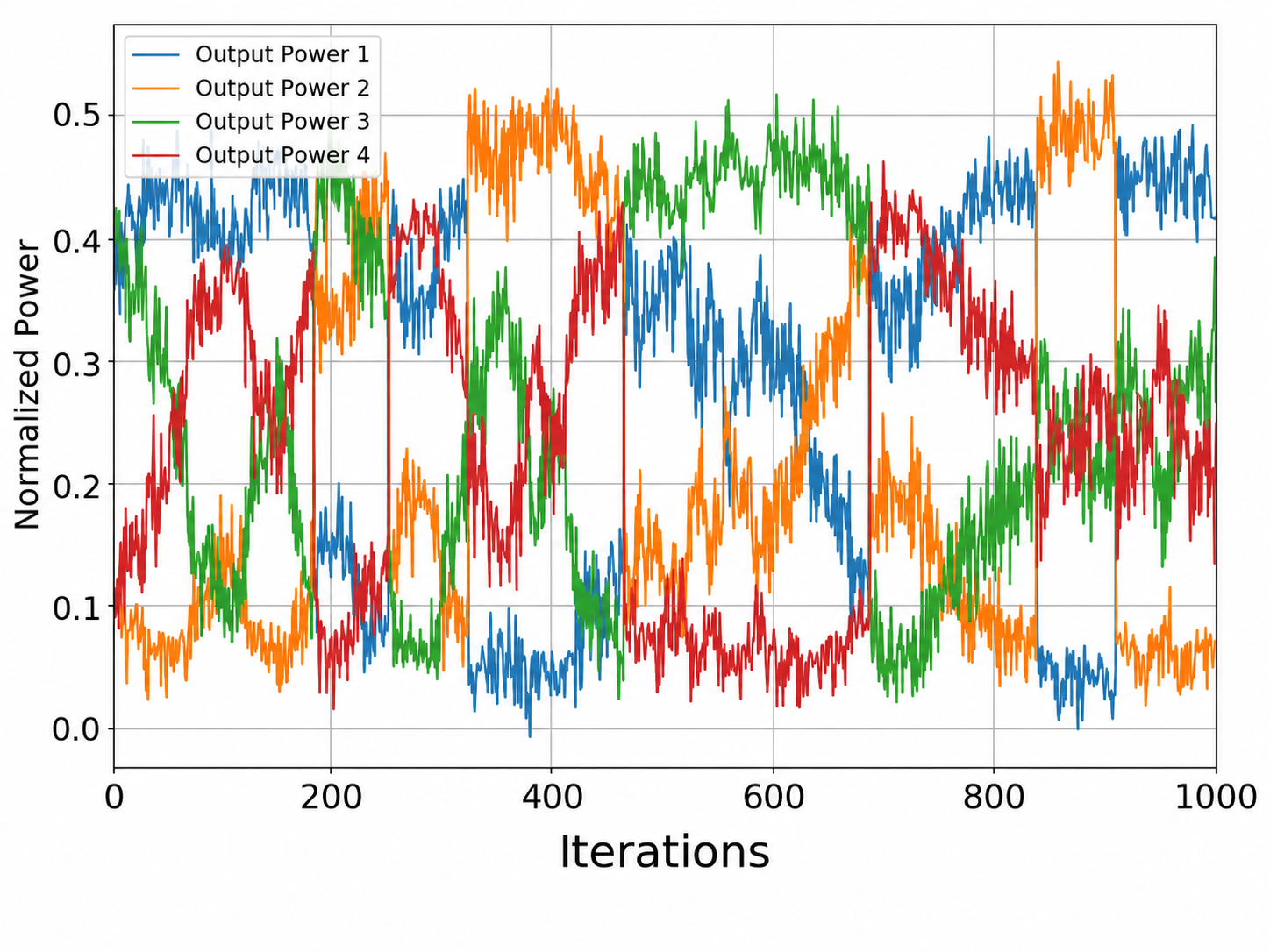}
    \caption{Measured optical output powers corresponding to Hadamard-2 decoding. Phase shifts of \(\pi\) were applied at known indices to simulate codeword transitions.}
    \label{fig:ploth2}
\end{figure}

Figure~\ref{fig:ploth2} shows the measured output powers for a test sequence of Hadamard-2 codewords. At randomly selected time indices, a relative \(\pi\)-phase shift was manually applied, leading to visible discontinuities in the output traces. In this experimental trace, transitions were identified at all seven applied shift locations: indices 186, 254, 326, 466, 688, 838, and 910.

Absolute codeword labels depend on the initial reference. For two-mode
signaling, pilot-free differential action-based encoding can instead
represent a bit by the presence or absence of a $\pi$ phase shift
between known symbol intervals. This identifies a transition without
requiring an absolute initial codeword label, but still depends on
inter-interval observations and timing information.

\section{Experimental Realization: Hadamard-4 Decoder}

The Hadamard-4 decoder was experimentally implemented using commercial programmable photonic processors 
(QuiX Quantum) fabricated on a low-loss silicon nitride platform \cite{chip2,bogaerts}. 
Each device consists of a universal multiport interferometer network with thermo-optic phase shifters 
that allow full reconfiguration of the implemented unitary. 
In our experiments, one photonic processor was programmed to prepare and encode the Hadamard codewords, 
while a second processor was configured to realize the eight-dimensional decoder transformation. 
The universal interferometer implements arbitrary unitary operations via the Clements decomposition \cite{clements2016}, 
which we used to compile the custom $8 \times 8$ circuit required for this work. 
The transmitter prepared four-mode Hadamard codewords, while the receiver implemented the distinct eight-dimensional unitary of Eq.~\eqref{eq:h8_unitary}; the decoding chip did not simply implement a second four-mode Hadamard transformation. 

The trigonometric variables are calculated from the normalized measured powers $P_1,\ldots,P_8$ using Eq.~\eqref{eq:trig_expressions} and the complete set of relations in the Supplementary Material. The inversion is exact for the programmed ideal transformation and equal-amplitude, four-mode input; residual gain imbalance or deviations from $U$ can perturb it. Output-power normalization removes common overall scaling, not unequal channel gains.

In the analytical Hadamard-4 experiment, the transmitter processor implemented the four-operation alphabet of Eq.~\eqref{eq:differential_alphabet}. At each scheduled symbol interval it either kept the three non-reference phases unchanged or applied $\pi$ shifts to one of the pairs $(B,C)$, $(B,D)$, and $(C,D)$. The selected operation, rather than the resulting absolute codeword, defined the transmitted symbol. The prescribed operation sequence supplied the ground truth independently of receiver decisions. A second processor held the fixed $8\times8$ decoding transformation throughout acquisition without optical phase-feedback updates. The eight measured output powers were normalized to unit total power and converted into six trigonometric observables using the reconstruction relations given in the Supplementary Material.

The initial measured observable-sign pattern established a temporal reference; knowledge of the initial \emph{absolute} Hadamard codeword label was not required for analytical operation identification. For each subsequent symbol interval, the receiver compared the new sign pattern with the preceding measured reference and identified which of the four allowed operations had been applied. The algorithm prioritized the three reconstructed observables with the largest absolute magnitudes to reduce sensitivity to zero crossings. The decoded output was the phase-change operation, with the reference updated for subsequent comparisons; if an initial codeword label is separately available, the absolute state can also be tracked by accumulating the detected operations. The analytical decoder achieved an overall classification accuracy of \textbf{96.19\%} on the measured sequence. This is differential decoding from successive measurements, not memoryless decoding without a reference, and does not establish tolerance to arbitrarily fast drift. Figure~\ref{fig:ploth4} shows the powers and reconstructed observables around the commanded updates.


To investigate the sensitivity of the analytical decoder to phase fluctuations, we performed numerical simulations using a Wiener phase-diffusion model, following the stochastic modeling framework investigated in our earlier study of phase instability in photonic processors~\cite{Elmas2025}. The phase evolution is represented by
\begin{equation}
\phi_{k+1}=\phi_k+\sigma_{\phi}\sqrt{\Delta t}\,Z_k,
\end{equation}
where $Z_k\sim\mathcal N(0,1)$, $\Delta t$ denotes the simulation time step, and $\sigma_{\phi}$ determines the phase-diffusion strength, with units of $\mathrm{s^{-1/2}}$ when $\Delta t$ is expressed in seconds.

\begin{figure}[!t]
\centering
\includegraphics[width=1\linewidth]{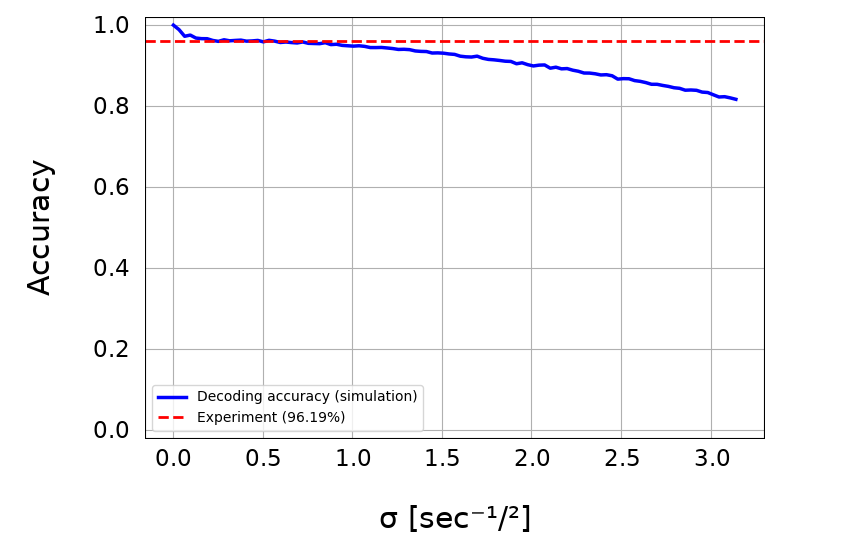}
\caption{Simulated analytical decoding accuracy versus phase-diffusion strength $\sigma_{\phi}$ (labeled $\sigma$ on the horizontal axis). The dashed line indicates the experimentally measured accuracy of 96.19\% for comparison.}
\label{fig:accuracy_sigma}
\end{figure}

\begin{figure}[!t]
    \centering
    \begin{subfigure}[b]{\columnwidth}
        \begin{overpic}[width=0.96\linewidth]{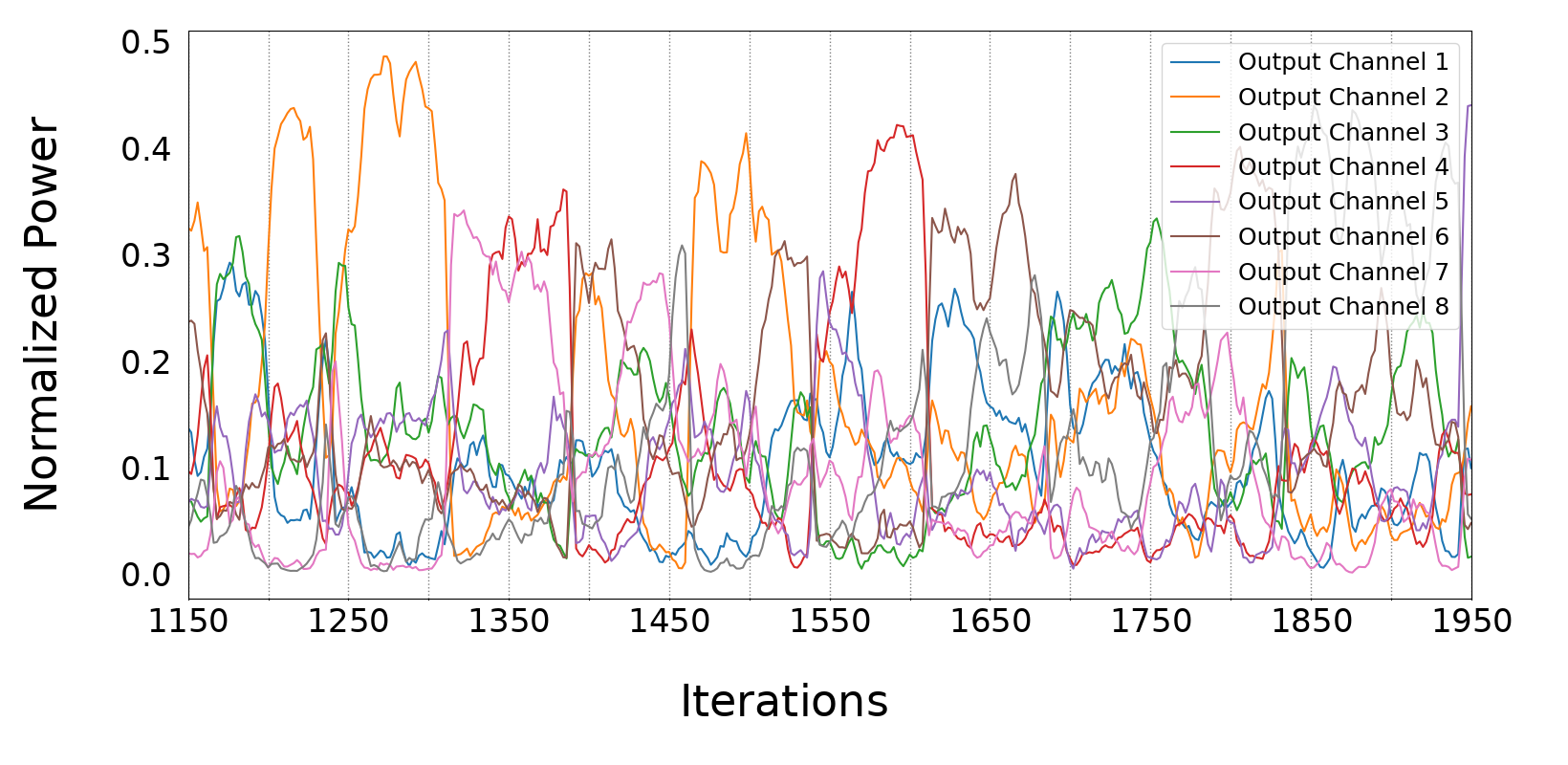}
            \put(0,50){\colorbox{white}{(a)}}
        \end{overpic}
        \vspace{2mm}
        \label{fig:ploth4a}
    \end{subfigure}
    \\
    
    \begin{subfigure}[b]{\columnwidth}
        \begin{overpic}[width=0.96\linewidth]{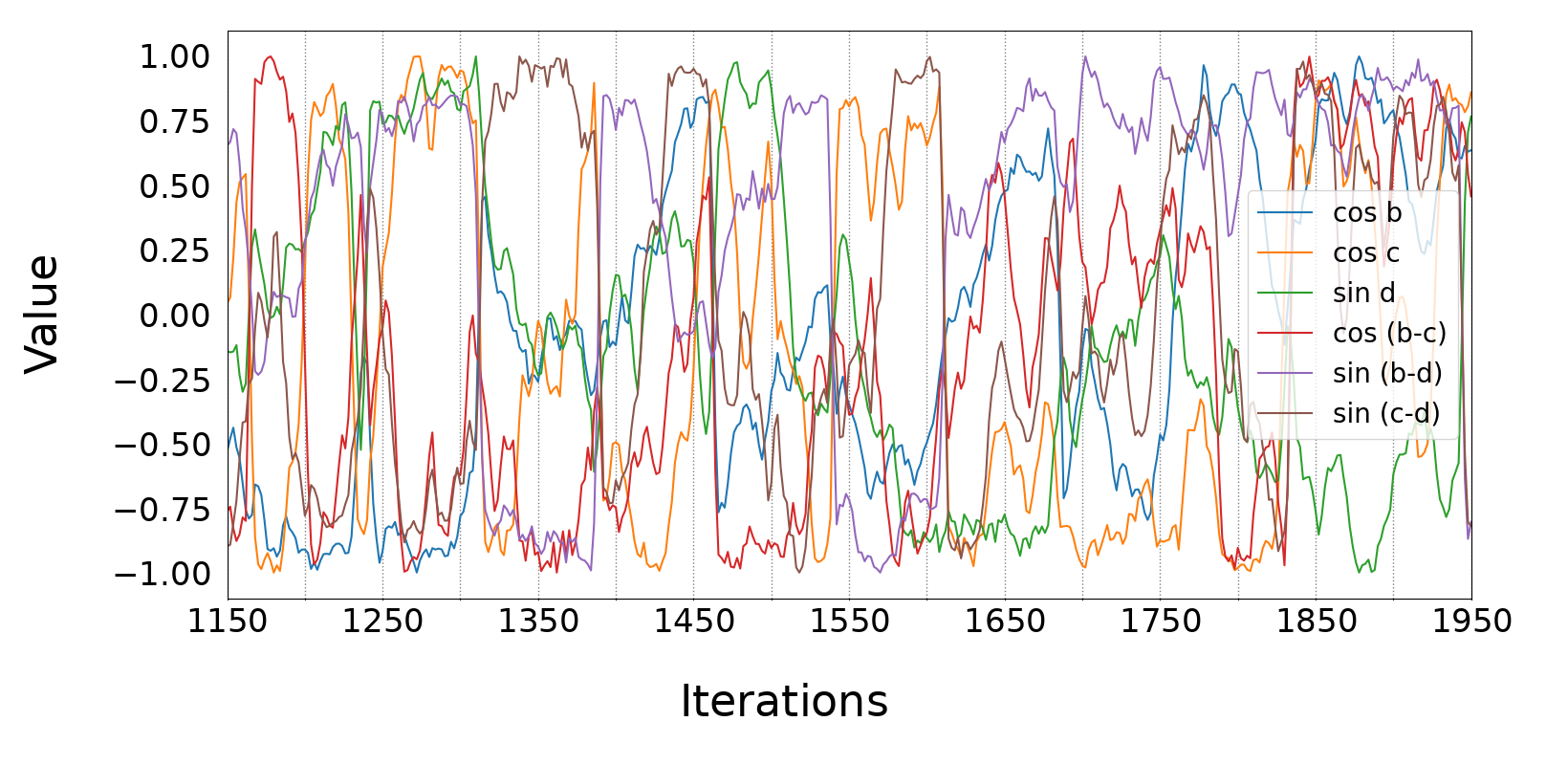}
            \put(0,48){\colorbox{white}{(b)}}
        \end{overpic}
        \label{fig:ploth4b}
    \end{subfigure}
    
    \caption{(a) Power distribution dynamics across 8 channels during codeword updates.
    (b) Reconstructed trigonometric observables used for differential transition identification.}
    \label{fig:ploth4}
\end{figure}

Figure~\ref{fig:accuracy_sigma} shows the simulated decoding accuracy as the phase-diffusion strength is varied. The dashed horizontal line indicates the separately measured analytical decoding accuracy of 96.19\% for comparison. The curve illustrates sensitivity to phase fluctuations under the assumed simulation conditions; it is not a fit of the diffusion coefficient to the experimental data.


The decoder is sensitive to phase changes that obscure the expected
sign transformations, particularly near zero crossings. Fabrication
and alignment imperfections, detector noise, and deviations from the
programmed transformation can also perturb reconstructed observables.
These limitations restrict the operating regime; they are not removed
by the sign-pattern approach.

\vspace{12pt}

Reliable operation identification requires that phase drift and measurement noise not obscure the expected sign transformations between observations; near-zero observables are especially susceptible to incorrect sign assignments.

\begin{figure}[!t]
    \centering
    \includegraphics[width=0.8\linewidth]{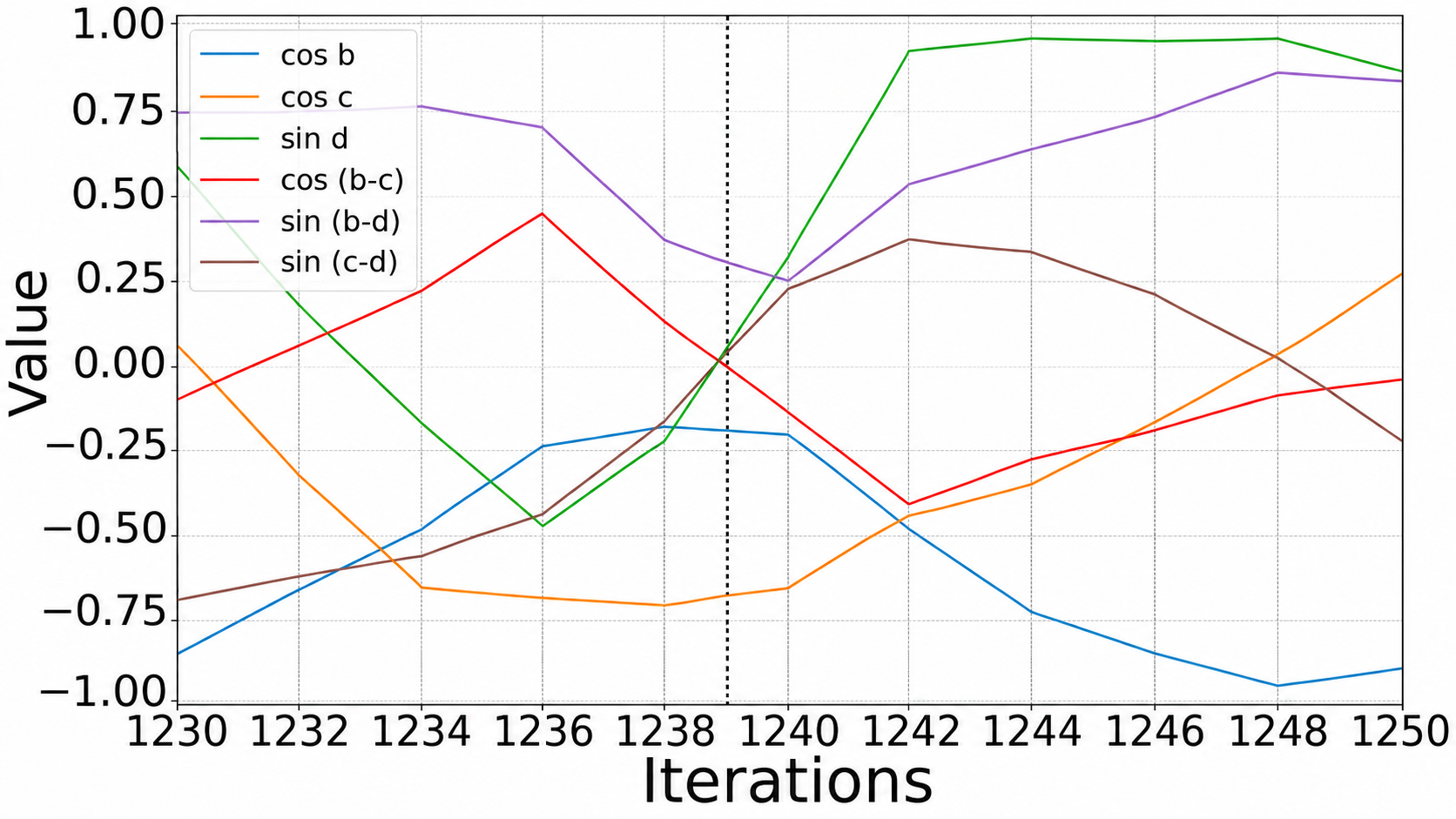}
    \caption{Reconstructed trigonometric observables around an example of analytical transition-decoding failure. The highlighted transition occurs near index 1239.}
    \label{fig:failed}
\end{figure}

To improve decoding reliability in the presence of noise and near-zero crossings, our implementation prioritized the three trigonometric quantities with the largest absolute values at each step. Specifically, we tracked sign changes in \(\cos B\), \(\cos C\), \(\sin D\), \(\cos(B - C)\), \(\sin(B - D)\), and \(\sin(C - D)\), and selected the three most significant (in magnitude) to determine codeword transitions. For instance, in a representative transition from Codeword~2 to Codeword~3 at index~1239, the decoding logic predicts sign changes in both \(\cos B\) and \(\sin D\). However, as illustrated in Figure~\ref{fig:failed}, \(\sin D\) does not exhibit the expected sign inversion. This discrepancy is consistent with phase fluctuations, including Brownian-like drift, which introduces stochastic fluctuations in the optical phase and can mask sign transitions when a trigonometric component is near zero. Such cases underscore the importance of dynamic significance-based selection among trigonometric variables, as relying on low-magnitude terms increases susceptibility to noise-induced errors.

The acquisition hardware was not specifically optimized for
low-noise optical measurement. Finite resolution and electronic noise
can therefore corrupt reconstructed observables near zero crossings;
their contribution to the measured error rate was not independently
characterized.

To mitigate these issues, careful calibration of the photonic chip is essential \cite{miller,elmascal}. This includes fine-tuning the relative phases, balancing splitter ratios, and compensating for fabrication variances through thermal or electrical control elements. Additional coding or optical feedback could be explored for more demanding noise conditions, at the cost of added complexity. Here the optical receiver is held fixed during the reported decoding measurements, and no overall electrical-energy or throughput advantage has been quantified.
\section{Machine–learning decoding}
\label{sec:last}
Second method to  find correct codeword with no phase stabilization is the machine - learning decoding that relies on \emph{sequential} inference: the previously decoded symbol supplies a phase reference and only the differential jump 
\(
\Delta\mathbf P
\)
around each detected transition is fed to the classifier. A single transition is thus represented by the eight components of $\Delta\mathbf P$ plus the categorical flag "initial state" ($0$–$3$); the label to be learned is the succeeding "final state".

To train and evaluate our decoders we relied on three physics-based \emph{synthetic data sets} and two \emph{experimental data sets}.  
Each synthetic data set contains \num{20\,000} labeled symbol transitions and captures a distinct noise regime:  
(i) an ideal, noise-free processor;  
(ii) the same processor disturbed solely by Brownian phase noise; and  
(iii) the processor perturbed by both Brownian phase noise and the \(\approx15\,\%\) detector-gain noise measured on our hardware (see Fig. 2).  
We then recorded two back-to-back experimental traces comprising \num{1\,502} and \num{1\,363} transitions, respectively; we use the first for fine-tuning and keep the second strictly for blind-test evaluation.  
For every transition we normalise the raw powers such that \(\sum_{i=1}^{8} P_i = 1\), yielding a nine-dimensional feature vector \(\mathbf{x}\in\mathbb{R}^9\) (the eight normalised power differences plus the categorical previous state) and a four-class target label \(y\in\{0,1,2,3\}\).

Six independent learners were trained: XGBoost, LightGBM, CatBoost, Random-Forest, Extra-Trees and a fully connected neural network (layers $64\!-\!32\!-\!32\!-\!16\!-\!8\!-\!4$ with \textsc{ReLU} activations, Adam optimiser $\eta=2\cdot10^{-5}$, early stopping after 70 idle epochs).  For the tree ensembles we used \num{200} estimators, depth 8–9 and learning rates 0.02–0.03; Random- and Extra-Trees were later enlarged by 100 trees via warm-start.  Two curricula were explored.  In the first, the three simulations were shown successively for 200 epochs each and the models were then fine-tuned for 50 epochs on the experimental training set.  In the second, the simulations were concatenated and shuffled once, processed for 300 epochs, and again fine-tuned for 50 epochs on the experiment.  Both delivered statistically identical results, so the joint curriculum—faster by \SI{18}{\%} wall time—was adopted.

\begin{figure}[!t]
  \centering
  \includegraphics[width=0.8\linewidth]{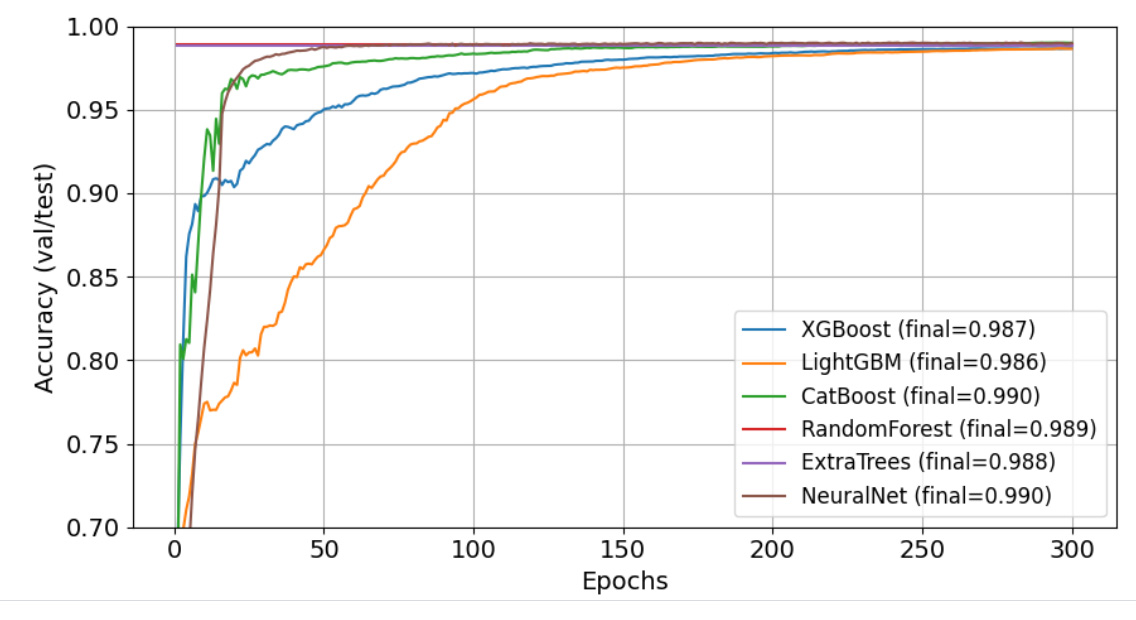}
  \caption{Validation accuracy on the mixed synthetic corpus; numbers in the legend are the final values.}
  \label{fig:epochs1}
\end{figure}

Figure\,\ref{fig:epochs1} plots the validation accuracy versus epoch for all six learners.  CatBoost and the neural net cross the $99\,\%$ line within forty epochs; XGBoost follows closely, while LightGBM climbs more gradually.  The shaded band (not shown to avoid clutter) is narrower than $\pm0.004$ for five random splits, confirming good statistical robustness.
\begin{table}[!t]
  \centering
  \caption{Top-1 accuracy of each learner after curriculum training.  Synthetic column refers to a held-out 10\,\% split; experimental column is the blind test.}
  \label{tab:acc}
  \begin{tabular}{lcc}
    \toprule
    \textbf{Model} & \textbf{Synthetic} & \textbf{Experimental}\\
    \midrule
    XGBoost       & 0.9866 & 0.9817\\
    LightGBM      & 0.9865 & 0.9685\\
    CatBoost      & 0.9900 & 0.9442\\
    Random-Forest & 0.9888 & 0.9560\\
    Extra-Trees   & 0.9879 & 0.9751\\
    Neural-Net    & 0.9902 & 0.9340\\
    \midrule
    \textbf{Soft-vote ensemble} & \textbf{0.9906} & \textbf{0.9831}\\
    \bottomrule
  \end{tabular}
\end{table}

\begin{table}[!t]
  \centering
  \caption{Confusion-matrix metrics for the ensemble on the blind experimental test set.}
  \label{tab:report}
  \begin{tabular}{lrrrr}
    \toprule
    \textbf{Class} & \textbf{Precision} & \textbf{Recall} & \textbf{$F_1$} & \textbf{Support}\\
    \midrule
    0 & 0.98 & 0.98 & 0.98 & 334\\
    1 & 0.98 & 0.98 & 0.98 & 348\\
    2 & 0.99 & 0.99 & 0.99 & 351\\
    3 & 0.98 & 0.99 & 0.99 & 330\\
    \midrule
    \textbf{Macro avg} & 0.98 & 0.98 & 0.98 & 1\,363\\
    \bottomrule
  \end{tabular}
\end{table}

Table\,\ref{tab:acc} summarises the top-1 accuracies after curriculum training.  On a held-out synthetic fold the ensemble reaches $0.9906$, and on the blind experimental trace $0.9831$.  Precision, recall and $F_1$ remain balanced across the four classes (see Table \ref{tab:acc}), the weakest transition ($1\!\rightarrow\!2$) still scoring $F_1=0.97$.  A confusion-matrix inspection shows only nearest-neighbour errors, consistent with occasional phase slips of magnitude $\approx\pi/8$.

In conclusion, a compact six-model ensemble trained on a blend of physics-based simulations and a single 90-s experimental trace suppresses the raw symbol error by more than two orders of magnitude, achieving $>98\,\%$ accuracy in a strongly perturbed $8\times8$ photonic processor without any optical phase stabilization.  The approach is immediately applicable to larger meshes and promises a low-power, fully passive path towards high-throughput reconfigurable optical networks.

\section{Conclusion}

We developed and experimentally tested a differential Hadamard-4 signaling and decoding scheme in which each information symbol specifies no phase change or a $\pi$ shift of two non-reference phases. Thus the information is encoded in the selected operation, not the absolute optical state. The fixed eight-port receiver reconstructs six trigonometric observables from simultaneous intensity measurements and identifies these operations through their known sign transformations, without continuous optical phase feedback. The analytical rule requires an earlier measured signature but not the initial absolute codeword label. The Supplementary Material establishes exact reconstruction, almost-everywhere distinction of the four operations apart from 24 exceptional phase configurations, and noise bounds conditional on a stable reference.

The analytical decoder identified the transmitted phase-change operations with 96.19\% accuracy on its measured sequence. A separate machine-learning analysis reports 98.31\% on held-out experimental data; these evaluations are not a matched head-to-head comparison. Neither result establishes memoryless decoding without a preceding observation or unrestricted tolerance to rapid phase diffusion.

We also compared the conventional four-port Hadamard decoder and the proposed eight-port receiver under an explicitly idealized, equal-detected-photon model. The added outputs resolve an example phase ambiguity of the four-port receiver and reduce conditional classification errors in the stated numerical comparison. These results identify a useful design trade-off; they are not a guarantee of superiority under arbitrary optical loss, detector noise, or drifting references.

Potential directions include structured phase-coded classical links, optical code-division multiple access, and phase-sensitive preprocessing in programmable photonic circuits. The phase-sensitive interference principle may also extend to other structured codebooks, including mutually unbiased bases~\cite{durt2010}, and connects to broader work in integrated optical neural networks and machine-learning-assisted photonics~\cite{Shen2017,Kudyshev2020}; any such implementation would require separate receiver-specific analyses. The fixed transformation avoids continuous input-phase feedback, but maintaining thermo-optic biases is not electrically free. Actual power consumption, latency, scalability, and network-level performance remain to be measured.

\section*{Funding}

This work was financed by the DFG via grant NO 1129/2-1
and by the Federal Ministry of Education and Research of
Germany via grants 16KIS1598K, 16KISQ093, 16KISQ077 and
16KIS2604. The generous support of the state of Bavaria
via the 6GQT project is greatly appreciated. Finally,
the authors acknowledge the financial support by the
Federal Ministry of Education and Research of Germany
in the programme of ``Souverän. Digital. Vernetzt.''.
Joint project 6G-life, project identification number:
16KISK002.

\section*{Disclosures}
The authors declare no conflicts of interest.

\section*{Data availability}
Data underlying the results presented in this paper are available from the
corresponding author upon reasonable request.

\clearpage
\section*{Supplementary Material}
\setcounter{section}{0}
\renewcommand{\thesection}{S\Roman{section}}
\setcounter{equation}{0}
\renewcommand{\theequation}{S\arabic{equation}}

\section{Detailed Conventional Four-Port Transition Calculation}
\subsection{Example: Codeword Flip and Small Noise}
\label{sup:subsec:H4-noise}

We illustrate the effect of phase noise on four-port Hadamard decoding by
considering a transition between two Hadamard codewords. Such a transition
changes the signs of exactly two components of the Hadamard row, corresponding
to $\pi$-phase shifts on two populated input modes. Here we consider a flip
of modes~2 and~3, corresponding to the relative phases $B$ and $C$.

Including additional phase perturbations $\Delta B$, $\Delta C$, and $\Delta D$
between the reference measurement and the new measurement, the output power
at port~1 is
\begin{align*}
P_{\mathrm{new},1}=&\frac{1}{8}\Big(2-\cos(B{+}\Delta B)\\
&-\cos(C{+}\Delta C)+\cos(D{+}\Delta D)\\
&+\cos[(B{+}\Delta B)-(C{+}\Delta C)]\\
&-\cos[(B{+}\Delta B)-(D{+}\Delta D)]\\
&-\cos[(C{+}\Delta C)-(D{+}\Delta D)]\Big).
\end{align*}
The expressions for the remaining output ports follow analogously.

The permutation of the output powers follows directly from the algebraic
structure of the Hadamard transformation, without requiring a small-phase
approximation. Define the input sign-flip operator \(D_{BC}=\operatorname{diag}(1,-1,-1,1)\).
For the normalized, Sylvester-ordered Hadamard matrix $H_4$ defined in
Section~\ref{sec:idealH4}, direct multiplication gives
\begin{equation}
H_4D_{BC}=\Pi_{BC}H_4,
\qquad
\Pi_{BC}=\begin{pmatrix}
0&0&0&1\\
0&0&1&0\\
0&1&0&0\\
1&0&0&0
\end{pmatrix}.
\label{eq:H4_BC_permutation_identity}
\end{equation}
Consequently, if $\mathbf y=H_4\mathbf x$ denotes the output before the
codeword transition, then in the absence of any additional drift
\(\mathbf y_{\mathrm{new}}=H_4D_{BC}\mathbf x=\Pi_{BC}\mathbf y\).
Taking squared magnitudes shows that the output powers are \emph{exactly}
permuted:
\begin{equation}
\begin{aligned}
P_{\mathrm{new},1}&=P_4, &
P_{\mathrm{new},2}&=P_3,\\
P_{\mathrm{new},3}&=P_2, &
P_{\mathrm{new},4}&=P_1.
\end{aligned}
\label{eq:H4_BC_power_permutation}
\end{equation}
This identity holds for arbitrary initial relative phases $B,C,D$; it is
not restricted to phases close to $0$ or $\pi$.

When further phase changes occur between measurements, let
$\boldsymbol{\Delta}=(\Delta B,\Delta C,\Delta D)^T$ and define the port
permutation by $\pi(1)=4$, $\pi(2)=3$, $\pi(3)=2$, and $\pi(4)=1$.
Because the intended input sign flips and the additional phase shifts
are diagonal operations, the measured powers obey
\begin{equation}
\begin{aligned}
P_{\mathrm{new},j}
&=P_{\pi(j)}(B+\Delta B,C+\Delta C,D+\Delta D)\\
&=P_{\pi(j)}(B,C,D)+O(\|\boldsymbol{\Delta}\|_2),
\end{aligned}
\label{eq:H4_BC_power_perturbed}
\end{equation}
where $P_j(B,C,D)$ denotes the pre-transition four-port power function.
Thus, the permutation itself is an exact consequence of Hadamard algebra;
only comparison with powers measured at an earlier phase configuration
becomes approximate when there is additional drift. This approximation
requires small phase changes during the comparison interval and does not
provide a guarantee under arbitrarily fast phase fluctuations.

\section{Minimax Optimality of the Quadrature Observables}

The reconstruction of both $\cos\alpha$ and $\sin\alpha$ is not an
arbitrary choice of phase-sensitive features. It is the minimax-optimal
choice among all pairs of equally normalized sinusoidal observables for
detecting an unknown-phase $\pi$ transition.

To see this, consider two equally populated optical input modes with
relative phase $\alpha$. For an arbitrary passive linear interferometer,
the power measured at output port $j$ has the general form
\(P_j(\alpha)= \frac{1}{2} \left| u_{j1}+u_{j2}e^{i\alpha} \right|^2=A_j+R_j\cos(\alpha-\beta_j)\),
where $A_j$ is independent of $\alpha$, $R_j\geq 0$ is the interference
amplitude, and $\beta_j$ is fixed by the interferometer. Thus, every
phase-dependent intensity observable obtainable from two populated
modes is necessarily a shifted and scaled sinusoid.

Consider now two equally normalized phase observables separated by an
arbitrary offset $\delta$, \(q_1(\alpha)=\cos\alpha\) and
\(q_2(\alpha)=\cos(\alpha-\delta)\).
A transition between the two Hadamard-2 codewords corresponds to
\(\alpha\longrightarrow\alpha+\pi\).
The resulting observable jump is
\begin{equation}
\Delta\mathbf{q}_{\delta}(\alpha)=-2\bigl(\cos\alpha,\cos(\alpha-\delta)\bigr)^T,
\end{equation}
and its squared Euclidean magnitude is
\begin{align}
\left\|
\Delta\mathbf{q}_{\delta}(\alpha)
\right\|_2^2
&=
4\left[
\cos^2\alpha+\cos^2(\alpha-\delta)
\right]
\\
&=
4\left[
1+\cos\delta\cos(2\alpha-\delta)
\right].
\end{align}
For a fixed separation $\delta$, the smallest transition signal over
all unknown initial phases is therefore
\(\min_{\alpha}\|\Delta\mathbf{q}_{\delta}(\alpha)\|_2^2=4(1-|\cos\delta|)\).
This worst-case value is maximized when
\(\delta=\frac{\pi}{2}\pmod{\pi}\), which corresponds exactly to the sine--cosine pair. At this optimum,
\(\Delta\mathbf{q}_{\pi/2}(\alpha)=-2(\cos\alpha,\sin\alpha)^T\),
and hence \(\left\|\Delta\mathbf{q}_{\pi/2}(\alpha)\right\|_2=2\)
for every initial phase $\alpha$.

The quadrature representation therefore has no blind initial phase.
Whenever one component is close to zero, the complementary component
remains finite. Indeed,
\(\max\left(|\cos\alpha|,|\sin\alpha|\right)\geq\frac{1}{\sqrt{2}}\),
so at least one reconstructed observable changes in magnitude by at least
$\sqrt{2}$ under every $\pi$ transition.

For the four-output transformation used here,
\(P_{1,2}(\alpha)=\frac{1}{4}\left(1\pm\cos\alpha\right)\),
\(P_{3,4}(\alpha)=\frac{1}{4}\left(1\pm\sin\alpha\right)\),
and the minimax-optimal observables are reconstructed directly as
\(\cos\alpha=2(P_1-P_2)\), \(\sin\alpha=2(P_3-P_4)\).
Moving from the raw powers to these quadrature observables therefore
converts the detector outputs into a phase-independent transition
representation whose total jump magnitude is constant and whose
strongest component is maximized in the worst case.

This optimality statement applies to pairs of equally normalized
sinusoidal observables. It does not assert global optimality among all
possible passive measurements with arbitrary numbers of output ports.
Different multi-output constructions may provide larger raw-power
separation under particular detector-noise models. The sine--cosine
construction is optimal specifically in the minimax sense relevant
here: among two-observable phase representations, it maximizes the
smallest detectable response to an unknown-phase $\pi$ transition.

\section{Phase-noise reference scales and relative-phase correlations}
\label{sec:ao_noise_reference}

Our previous experimental study of phase instability in an eight-mode photonic
processor~\cite{Elmas2025} used Brownian phase fluctuations as a simplified
model for fiber-induced drift. The phase-distribution comparison used a
Brownian coefficient $\sigma=0.25\,\mathrm{s^{-1/2}}$, with 35 experimental
runs of 300~s and approximately 125 recorded output-power samples per second
per channel. A separate on-chip feedback comparison used
$\sigma=0.15\,\mathrm{s^{-1/2}}$ with a 38-Hz feedback update rate.
These values characterize the earlier experiments and provide reference
scales for the phase-diffusion model. They are not measurements of the phase
diffusion in the present Hadamard-4 decoding run, and neither the sampling
rate nor the 38-Hz feedback update rate establishes the numerical time step
used for Fig.~3 of the main paper. The actual generator time step, detector
noise level, and decision count in that figure must be documented from its
original simulation settings.

There is an additional distinction between independent input-mode phase
walks and the three relative phases used by the Hadamard-4 decoder. If
four absolute input phases $\theta_j$ follow independent Brownian motions
$d\theta_j=\sigma_{\mathrm{mode}}\,dW_j$, and
$\boldsymbol\beta=(\theta_2-\theta_1,\theta_3-\theta_1,
\theta_4-\theta_1)^T$, then their increments satisfy
\begin{equation}
\operatorname{Cov}(\Delta\boldsymbol\beta)
=\sigma_{\mathrm{mode}}^2\Delta t\,
\begin{pmatrix}2&1&1\\1&2&1\\1&1&2\end{pmatrix}.
\end{equation}
Consequently, the standard deviation of each relative-phase increment is
$\sqrt{2}\,\sigma_{\mathrm{mode}}\sqrt{\Delta t}$ and distinct relative
increments have correlation $1/2$. A simulation which assigns independent
Wiener increments directly to $B,C,D$ is therefore not identical to the
independent-input-mode model. The earlier reported Brownian coefficients
cannot be transferred between these conventions without specifying which
phase variable they describe.

\section{Generalization Beyond Hadamard Transitions}
\label{sec:generalization}

The principle that a passive interferometer maps phase differences into intensity-dependent interference terms is not restricted to Hadamard inputs. However, exact recovery, distinguishability, and phase-noise robustness do not automatically extend to an arbitrary codebook; they depend on the receiver and the available measurements. To state the generic algebraic principle, consider a family of mutually orthogonal equal-amplitude phase codewords. Let each codeword be written as
\begin{equation}
    \ket{\psi^{(m)}} = \frac{1}{\sqrt{N}}
    \left( e^{i\phi^{(m)}_1}, e^{i\phi^{(m)}_2}, \ldots, e^{i\phi^{(m)}_N} \right)^T,
\end{equation}
with orthogonality condition
\begin{equation}
    \langle\psi^{(m)}\mid\psi^{(n)}\rangle = 
    \frac{1}{N}\sum_{k=1}^N 
    e^{i(\phi^{(n)}_k - \phi^{(m)}_k)} = 0,
    \qquad m\neq n.
\end{equation}

When such a codeword passes through a passive interferometer $U$, the output field at port $j$ is
\((U\psi^{(m)})_j = \frac{1}{\sqrt{N}}\sum_{k=1}^N U_{jk} e^{i\phi^{(m)}_k}\),
and the detected power is
\begin{equation}
    P_j^{(m)} 
    = |(U\psi^{(m)})_j|^2
    = \frac{1}{N} \sum_{a,b=1}^N 
    U_{ja}U_{jb}^*
    e^{i(\phi^{(m)}_a - \phi^{(m)}_b)}.
\end{equation}
The diagonal terms $a=b$ contribute a constant offset
\(A_j = \frac{1}{N}\sum_{k=1}^N |U_{jk}|^2\), while all phase dependence arises from the off-diagonal interference terms.

A general unitary interferometer contains complex-valued coefficients, so each product $U_{ja}U_{jb}^*$ can be written as
\(U_{ja}U_{jb}^* = \alpha_{j,ab} + i\beta_{j,ab}\), with
\(\alpha_{j,ab},\beta_{j,ab}\in\mathbb{R}\).
Expressing the input phase difference as
\(e^{i(\phi^{(m)}_a-\phi^{(m)}_b)}=\cos\Delta^{(m)}_{ab}+i\sin\Delta^{(m)}_{ab}\),
\(\Delta^{(m)}_{ab}=\phi^{(m)}_a-\phi^{(m)}_b\),
shows that the real-valued intensity measurement necessarily involves both trigonometric quadratures:
\begin{align}
    \Re\!\left[
        U_{ja}U_{jb}^* e^{i\Delta^{(m)}_{ab}}
    \right]
    = 
    \alpha_{j,ab}\cos\Delta^{(m)}_{ab}
    - 
    \beta_{j,ab}\sin\Delta^{(m)}_{ab}.
\end{align}
Consequently,
\begin{equation}
    P_j^{(m)} = A_j
    + \sum_{a<b}
    \left[
        c_{j,ab}\cos\Delta^{(m)}_{ab}
        + s_{j,ab}\sin\Delta^{(m)}_{ab}
    \right],
\end{equation}
with coefficients \(c_{j,ab} = \frac{2}{N}\alpha_{j,ab}\) and
\(s_{j,ab} = -\frac{2}{N}\beta_{j,ab}\), consistent with the definitions in Eqs. \eqref{eq:pow1} and \eqref{eq:Palpha}.
Both cosine and sine contributions can arise when the interferometer coupling coefficients have appropriate real and imaginary components; their individual observability depends on the chosen transformation. The real part of the coupling contributes to the cosine channel, while the imaginary part contributes to the sine channel.

Each phase codeword can be associated, up to a global phase, with a trigonometric signature
\(\mathbf{t}^{(m)}=\left(\cos\Delta^{(m)}_{ab},\sin\Delta^{(m)}_{ab}\right)_{1\leq a<b\leq N}\),
and transitions between codewords correspond to deterministic transformations of these signatures. In Hadamard-encoded systems these appear as sign flips in $\cos\Delta$, whereas for general codes they involve simultaneous changes in multiple cosine and sine components. The structure, however, remains fully deterministic.

In the presence of small phase noise $\delta_k$ on each mode, the differences shift as
\(
\Delta^{(m)}_{ab}\rightarrow \Delta^{(m)}_{ab}+\delta_{ab}
\)
with $\delta_{ab}=\delta_a-\delta_b$. First-order expansions give
\(\cos(\Delta+\delta)\approx\cos\Delta-\delta\sin\Delta\),
\(\sin(\Delta+\delta)\approx\sin\Delta+\delta\cos\Delta\),
illustrating that the two quadratures exhibit complementary noise responses. Whenever one observable becomes unreliable due to a zero crossing, the other typically remains strong, providing intrinsic redundancy that enhances robustness.

These expressions establish the generic possibility of sinusoidal intensity responses, not generic access to every quadrature or an invertible measurement map. For other codebooks, suitably chosen interferometers may yield useful distinguishable signatures, but this requires an independent identifiability and noise analysis. Mutually unbiased bases are one possible structured family of interest~\cite{durt2010}; no such extension is implemented or proven here. Related hardware-oriented contexts include integrated optical neural networks~\cite{Shen2017} and machine-learning-assisted integrated photonics~\cite{Kudyshev2020}.

\section{Algebraic Identifiability and Noise Robustness of Hadamard Codeword Transitions}
\label{app:identifiability}

\subsection{Exact reconstruction from output powers}
\label{app:reconstruction}

We first verify that the six trigonometric observables used by the decoder
are directly reconstructible from the eight output-power measurements.
With the input state in Eq.~\eqref{eq:h8_input} and the fixed unitary transformation in Eq.~\eqref{eq:h8_unitary},
let
\(\mathbf p=(P_1,\ldots,P_8)^T\),
\(P_j=|(U\mathbf x_{\mathrm{in}})_j|^2\),
\(\sum_{j=1}^8 P_j=1\).
Define
\begin{equation}
\begin{aligned}
\mathbf q(B,C,D)=\bigl(&\cos B,\cos C,\sin D,\\
&\cos(B-C),\sin(B-D),\sin(C-D)\bigr)^T.
\end{aligned}
\label{eq:app_observables}
\end{equation}
Direct matrix multiplication gives the affine mapping
\begin{equation}
\mathbf p=\frac18\mathbf 1_8+\frac1{16}A\mathbf q,
\label{eq:app_power_mapping}
\end{equation}
where
\begin{equation}
A=\begin{pmatrix}
-1&-1&-1& 1&-1&-1\\
-1& 1& 1&-1& 1&-1\\
-1&-1& 1& 1& 1& 1\\
-1& 1&-1&-1&-1& 1\\
 1& 1& 1& 1&-1&-1\\
 1&-1&-1&-1& 1&-1\\
 1&-1& 1&-1&-1& 1\\
 1& 1&-1& 1& 1& 1
\end{pmatrix}.
\label{eq:app_mapping_matrix}
\end{equation}
The columns of $A$ are mutually orthogonal:
\(A^TA=8I_6\) and \(A^T\mathbf 1_8=\mathbf 0\).
Hence the exact inverse formula is
\begin{equation}
\mathbf q=2A^T\mathbf p.
\label{eq:app_exact_inverse}
\end{equation}
For normalized measured powers, this is equivalently
\begin{align}
\cos B &=4(P_5+P_6+P_7+P_8)-2,\label{eq:app_cosB}\\
\cos C &=4(P_2+P_4+P_5+P_8)-2,\label{eq:app_cosC}\\
\sin D &=4(P_2+P_3+P_5+P_7)-2,\label{eq:app_sinD}\\
\cos(B-C)&=4(P_1+P_3+P_5+P_8)-2,\label{eq:app_cosBC}\\
\sin(B-D)&=4(P_2+P_3+P_6+P_8)-2,\label{eq:app_sinBD}\\
\sin(C-D)&=4(P_3+P_4+P_7+P_8)-2.\label{eq:app_sinCD}
\end{align}
These identities hold for arbitrary $B,C,D$, rather than only ideal
Hadamard phases. Thus all six observables can be recovered from intensity
measurements alone in the ideal model. The six quantities are constrained by the three input-relative phases and should not be interpreted as six independent degrees of freedom.

\subsection{Hadamard transition operators}

The phases $B,C,D$ are relative to the first populated input mode.
Information is encoded in the selected operation, not in the absolute
Hadamard codeword. At each symbol interval the transmitter chooses no change
or a $\pi$ shift on exactly two of the three non-reference phases:
\(\mathcal G=\{I,T_{BC},T_{BD},T_{CD}\}\).
These operations form a four-symbol alphabet independently of the initial
codeword label. For an unchanged pre-transition phase reference, they act on
the reconstructed observables through diagonal sign matrices,
$\mathbf q\mapsto S_g\mathbf q$, where
\begin{align}
S_I &=I_6,\\
S_{BC}&=\operatorname{diag}(-1,-1,1,1,-1,-1),\\
S_{BD}&=\operatorname{diag}(-1,1,-1,-1,1,-1),\\
S_{CD}&=\operatorname{diag}(1,-1,-1,-1,-1,1).
\end{align}
The matrices satisfy $S_g^2=I_6$ and $S_{BC}S_{BD}=S_{CD}$,
with corresponding cyclic identities. The transition group is therefore
isomorphic to $\mathbb Z_2\times\mathbb Z_2$.

\subsection{Almost-everywhere identifiability}

\noindent\textbf{Theorem A1 (Almost-everywhere identifiability).}
\emph{For every $(B,C,D)\in[0,2\pi)^3$ except a finite set,
the four candidate signatures $\mathbf q$, $S_{BC}\mathbf q$,
$S_{BD}\mathbf q$, and $S_{CD}\mathbf q$ are pairwise distinct.}

\medskip\noindent\emph{Proof.}
For any two candidates define
\(d_{g,h}^2=\|S_g\mathbf q-S_h\mathbf q\|_2^2=\|\mathbf q-S_gS_h\mathbf q\|_2^2\).
The second equality follows because the commuting sign matrices are
orthogonal and symmetric. Thus every pairwise distance is one of
\begin{align}
d_{BC}^2&=4\bigl[\cos^2 B+\cos^2 C
+\sin^2(B-D)+\sin^2(C-D)\bigr],\label{eq:app_dbc}\\
d_{BD}^2&=4\bigl[\cos^2 B+\sin^2 D
+\cos^2(B-C)+\sin^2(C-D)\bigr],\label{eq:app_dbd}\\
d_{CD}^2&=4\bigl[\cos^2 C+\sin^2 D
+\cos^2(B-C)+\sin^2(B-D)\bigr].\label{eq:app_dcd}
\end{align}
Each right-hand side is a sum of nonnegative terms and vanishes only
when every term vanishes. Their zero sets are
\begin{align}
d_{BC}=0&\iff B=C=D=\frac\pi2\pmod\pi,\\
d_{BD}=0&\iff B=\frac\pi2,\quad C=D=0\pmod\pi,\\
d_{CD}=0&\iff B=D=0,\quad C=\frac\pi2\pmod\pi.
\end{align}
Each family has eight members in $[0,2\pi)^3$, and the families
are disjoint. Therefore there are exactly 24 exceptional phase
configurations. At all other configurations every nonidentity
transition distance is positive, so all four candidates are distinct.
\hfill$\square$

\medskip\noindent\textbf{Corollary A1.}
If $(B,C,D)$ has a probability distribution absolutely continuous
with respect to Lebesgue measure, the probability of exact transition
ambiguity is zero.

This concerns identification of the transmitted operation from a known,
unchanged \emph{measured} reference signature. The initial absolute Hadamard
codeword label need not be known, but the preceding signature must be
available and stable over the comparison interval.

\subsection{Minimum signature separation}

Define
\(d_{\min}(B,C,D)=\min\{d_{BC},d_{BD},d_{CD}\}\).
Theorem A1 implies $d_{\min}>0$ outside the exceptional set, but
$d_{\min}$ approaches zero near that set. Thus the result does not
give a phase-uniform noise margin.

For a candidate transition $g$, Eq.~\eqref{eq:app_power_mapping}
gives the ideal power vector
\(\mathbf p_g=\frac18\mathbf 1_8+\frac1{16}AS_g\mathbf q\).
Using $A^TA=8I_6$,
\begin{equation}
\begin{aligned}
\|\mathbf p_g-\mathbf p_h\|_2^2
&=\frac1{256}\|A(S_g\mathbf q-S_h\mathbf q)\|_2^2\\
&=\frac1{32}\|S_g\mathbf q-S_h\mathbf q\|_2^2.
\end{aligned}
\label{eq:app_power_distance}
\end{equation}
Hence the mapping scales every squared pairwise signature distance
by $1/32$. Exact ambiguities in observable space and in ideal output
power space coincide.

\subsection{Ideal photon-counting comparison protocol}

The quantitative four-port/eight-port comparison in Section~\ref{subsec:controlled-comparison} is a numerical ideal-model benchmark, separate from either experimental decoding algorithm. The pre-transition phases are supplied without uncertainty to both receivers, each trial uses the same fixed number of detected photons $n$, and a conditional multinomial likelihood is maximized across the four possible transitions. Each random phase triple is drawn independently and uniformly on $[0,2\pi)^3$, with a uniformly random true transition. This benchmark excludes phase drift within the detection interval, loss and channel-dependent detector effects, and uncertainty in the pre-transition reference. The finite-sample estimates are illustrated in Fig.~\ref{fig:receiver_comparison}; reproducibility code and exact counts are included in the accompanying source archive.


\subsection{Conditional error bound under Gaussian noise}

Assume that the reference $\mathbf q$ is known and remains fixed during
the transition comparison, and model the reconstructed observation as
\begin{equation}
\mathbf y=S_g\mathbf q+\boldsymbol\eta,
\qquad \boldsymbol\eta\sim\mathcal N(0,\sigma_{\mathrm{obs}}^2I_6).
\label{eq:app_noise}
\end{equation}
Here $\sigma_{\mathrm{obs}}$ denotes noise in the reconstructed observables, not the Wiener phase-diffusion strength of the simulation in the main paper. Consider the minimum-distance decoder
\begin{equation}
\widehat g=\operatorname*{arg\,min}_{h\in\mathcal G}
\|\mathbf y-S_h\mathbf q\|_2^2.
\label{eq:app_rule}
\end{equation}

\noindent\textbf{Theorem A2 (Conditional Gaussian error bound).}
\emph{Under Eqs.~\eqref{eq:app_noise}--\eqref{eq:app_rule},}
\begin{equation}
\begin{aligned}
P_{\mathrm{error}}(B,C,D)\leq\min\Bigl\{1,\;\\
\tfrac32\exp\bigl[-d_{\min}^2(B,C,D)/(8\sigma_{\mathrm{obs}}^2)\bigr]\Bigr\}.
\end{aligned}
\label{eq:app_bound}
\end{equation}

\medskip\noindent\emph{Proof.}
For an incorrect hypothesis $h\ne g$, set
$\boldsymbol\Delta=S_h\mathbf q-S_g\mathbf q$.
The decoder can prefer $h$ only if
\(2\boldsymbol\eta^T\boldsymbol\Delta\geq\|\boldsymbol\Delta\|_2^2\).
Since $\boldsymbol\eta^T\boldsymbol\Delta$ is Gaussian with variance
$\sigma_{\mathrm{obs}}^2\|\boldsymbol\Delta\|_2^2$, the probability of this event is
\begin{equation}
Q\!\left(\frac{\|\boldsymbol\Delta\|_2}{2\sigma_{\mathrm{obs}}}\right)
\leq\tfrac12 e^{-\|\boldsymbol\Delta\|_2^2/(8\sigma_{\mathrm{obs}}^2)}.
\end{equation}
There are three incorrect hypotheses, and every pairwise distance is
at least $d_{\min}$. The union bound, combined with the trivial upper
bound of one, proves Eq.~\eqref{eq:app_bound}. \hfill$\square$

This bound applies to the ideal nearest-signature decoder with a known
reference. It does not directly bound a sign-only sequential decoder
under drift between measurements.

\subsection{Correlated observable noise}

The six observables are reconstructed from the same eight powers,
so their measurement errors can be correlated. Let
\(\widetilde{\mathbf p}=\mathbf p+\boldsymbol\xi\), with
\(\operatorname{Cov}(\boldsymbol\xi)=C_p\).
Applying the exact linear reconstruction produces
\(\widetilde{\mathbf q}=2A^T\widetilde{\mathbf p}=\mathbf q+2A^T\boldsymbol\xi\),
and consequently
\(\Sigma=4A^TC_pA\).
If $C_p=\sigma_P^2I_8$, then \(\Sigma=32\sigma_P^2I_6\).
Thus the isotropic observable-noise model corresponds to independent,
equal-variance additive noise on the eight \emph{unnormalized} powers. If the recorded powers are first divided by their fluctuating sum, the resulting normalization generally introduces inter-channel correlations, and $C_p$ must refer to the covariance after that processing step. Unequal detector variances and other noise mechanisms may also yield non-isotropic observable noise.

For Gaussian observable noise with fixed positive-definite covariance
$\Sigma$, the maximum-likelihood rule is
\begin{equation}
\widehat g=\operatorname*{arg\,min}_{h\in\mathcal G}
(\mathbf y-S_h\mathbf q)^T\Sigma^{-1}
(\mathbf y-S_h\mathbf q).
\end{equation}
Define
\begin{equation}
D_{\min}^2=\min_{g\ne h}
(S_g\mathbf q-S_h\mathbf q)^T\Sigma^{-1}
(S_g\mathbf q-S_h\mathbf q).
\end{equation}
The Gaussian pairwise-error argument yields
\(P_{\mathrm{error}}\leq\min\{1,\tfrac32\exp(-D_{\min}^2/8)\}\).

These results establish conditional distinguishability under explicit
measurement models. They do not assume that Brownian phase evolution
is negligible over arbitrarily long comparison intervals, and they
do not provide unconditional guarantees under arbitrary phase drift
or inaccurate reference estimation.


\begin{thebibliography}{99}
\bibitem{Li}
Li, W., Wang, D., and Liu, J., 2024." Robust Walsh–Hadamard Transform Based Imaging Method for Low-SNR Scenes." \textit{Sensors}, \textbf{24}(10), 3062. \url{https://doi.org/10.3390/s24103062}

\bibitem{Hadamard_CDMA}
Wei, Z. and Ghafouri-Shiraz, H., 2002. "Codes for spectral-amplitude-coding optical CDMA systems." \textit{Journal of lightwave technology}, \textbf{20}(8), p.1284.
\bibitem{Noshad}
Noshad, M. and Brandt-Pearce, M., 2016. "Hadamard-coded modulation for visible light communications." \textit{IEEE Transactions on Communications}, 64(3), pp.1167-1175.

\bibitem{Luo}Luo, W., Cao, L., Shi, Y., Wan, L., Zhang, H., Li, S., Chen, G., Li, Y., Li, S., Wang, Y. and Sun, S., 2023. "Recent progress in quantum photonic chips for quantum communication and internet.", \textit{Light: Science \& Applications}, \textbf{12}(1), p.175.
\bibitem{Flamini2019}
Flamini, F., Spagnolo, N. and Sciarrino, F., 2018. "Photonic quantum information processing: a review.",\textit{ Reports on Progress in Physics}, 82(1), p.016001.
\bibitem{aghaee}
Aghaee Rad, H., Ainsworth, T., Alexander, R.N., Altieri, B., Askarani, M.F., Baby, R., Banchi, L., Baragiola, B.Q., Bourassa, J.E., Chadwick, R.S. and Charania, I., 2025. "Scaling and networking a modular photonic quantum computer.", \textit{Nature}, 638(8052), pp.912-919.

\bibitem{maring}
Maring, N., Fyrillas, A., Pont, M., Ivanov, E., Stepanov, P., Margaria, N., Hease, W., Pishchagin, A., Lemaître, A., Sagnes, I. and Au, T.H., 2024. "A versatile single-photon-based quantum computing platform." \textit{Nature Photonics}, 18(6), pp.603-609.
\bibitem{Elmas2025}
Elmas, G., Litvin, I., Kohl, P., and Nötzel, J., 2025. "Modeling and analysis of phase instability in a photonic processor." \textit{Applied Optics}, 64(14), 3995-4003.

\bibitem{Litvin2025}
Litvin, I. A., Elmas, G., El-Safty, K., Chaudhary, S., and Nötzel, J. (2025). "Robust calibration and energy optimization in reconfigurable photonic processors." \textit{Optics Express}, 33(16), 35011-35027.


\bibitem{Litvin2}
I. A. Litvin, G. Elmas, P. Kohl and J. Nötzel, "Stable Signal Processing with a Photonic Processor," in \textit{Journal of Lightwave Technology}, doi: 10.1109/JLT.2025.3606027.



\bibitem{elmascal}
G. Elmas, I. A. Litvin, and J. Nötzel,
``Local Variance-Based Calibration of Programmable
Photonic Interferometer Meshes,''
\emph{Journal of Lightwave Technology},
submitted for publication, 2026.
[Online]. Available: https://arxiv.org/abs/2607.14759.

\bibitem{Madsen2022}
Madsen, L. S., Laudenbach, F., et al., 2022." A quantum computational advantage with a programmable photonic processor." \textit{Nature}, \textbf{606}(7912), 75–81. \url{https://doi.org/10.1038/s41586-022-04725-x}

\bibitem{Reck1994}
Reck, M., Zeilinger, A., Bernstein, H.J. and Bertani, P., 1994. "Experimental realization of any discrete unitary operator." \textit{Physical review letters}, 73(1), p.58.
\bibitem{Cui}


Cui, C., Postlewaite, J., Saif, B.N., Fan, L. and Guha, S., 2025. "Superadditive communication with the green machine as a practical demonstration of nonlocality without entanglement." \textit{Nature Communications}, \textbf{16}(1), p.3760.
\bibitem{Qiang2018}
Qiang, X., Zhou, X., Wang, J., Wilkes, C.M., Loke, T., O’Gara, S., Kling, L., Marshall, G.D., Santagati, R., Ralph, T.C. and Wang, J.B., 2018. "Large-scale silicon quantum photonics implementing arbitrary two-qubit processing." \textit{Nature photonics}, 12(9), pp.534-539.
\bibitem{Shen2017}
Shen, Y., Harris, N. C., Skirlo, S., et al., 2017. "Deep learning with coherent nanophotonic circuits." \textit{Nature Photonics}, \textbf{11}(7), 441–446.\url{https://doi.org/10.1038/nphoton.2017.93}
\bibitem{Harris}
Harris, N.C., Steinbrecher, G.R., Prabhu, M., Lahini, Y., Mower, J., Bunandar, D., Chen, C., Wong, F.N., Baehr-Jones, T., Hochberg, M. and Lloyd, S., 2017. "Quantum transport simulations in a programmable nanophotonic processor." \textit{Nature Photonics}, 11(7), pp.447-452.
\bibitem{Kudyshev2020}
Kudyshev, Z. A., Shalaev, V. M., and Boltasseva, A., 2021. "Machine Learning for Integrated Quantum Photonics." \textit{ACS Photonics}, \textbf{8}(1), 34--46. doi:10.1021/acsphotonics.0c00960.

\bibitem{chip1}
C. G. H. Roeloffzen et al., "Low-Loss Si3N4 TriPleX Optical Waveguides: Technology and Applications Overview," in \textit{IEEE Journal of Selected Topics in Quantum Electronics}, vol. 24, no. 4, pp. 1-21, July-Aug. 2018, Art no. 4400321, doi: 10.1109/JSTQE.2018.2793945.
\bibitem{chip2}
Caterina Taballione, Tom A. W. Wolterink, Jasleen Lugani, Andreas Eckstein, Bryn A. Bell, Robert Grootjans, Ilka Visscher, Dimitri Geskus, Chris G. H. Roeloffzen, Jelmer J. Renema, Ian A. Walmsley, Pepijn W. H. Pinkse, and Klaus-J. Boller, "8×8 reconfigurable quantum photonic processor based on silicon nitride waveguides," \textit{Opt. Express 27}, 26842-26857 (2019)

\bibitem{bogaerts}
Bogaerts, W., Pérez, D., Capmany, J., Miller, D. A., Poon, J., Englund, D., ... \& Melloni, A. (2020). "Programmable photonic circuits." \textit{Nature}, 586(7828), 207-216.
\bibitem{clements2016}
Clements, W. R., Humphreys, P. C., Metcalf, B. J., Kolthammer, W. S., \& Walmsley, I. A. (2016). "Optimal design for universal multiport interferometers." \textit{Optica}, 3(12), 1460-1465.

\bibitem{miller}
David A. B. Miller, "Perfect optics with imperfect components," \textit{Optica 2}, 747-750 (2015)


\bibitem{LitvinMultiInput2026}
I. A. Litvin, G. Elmas, P. Kohl, and J. N\"otzel,
``Multi-input signal phase stabilization in photonic processors with on-chip feedback control,''
\emph{Optics Express}, \textbf{34}(6), 11244--11258 (2026), doi:10.1364/OE.579969.

\bibitem{LitvinJointDetection2026}
I. A. Litvin, G. Elmas, and J. N\"otzel,
``Joint Detection on Reconfigurable Photonic Processors With Multi-Input Phase Stabilization,''
\emph{IEEE Photonics Journal} (2026), doi:10.1109/JPHOT.2026.3683204.

\bibitem{Paryzkova2026}
M. Par\'{y}zkov\'{a}, C. S. Hamilton, I. Jex, M. Stefszky, and C. Silberhorn,
``Input phase noise in Gaussian boson sampling,''
\emph{Physical Review A}, \textbf{114}, 022203 (2026), doi:10.1103/pddb-5z5q.

\bibitem{ElmasQuadrature2026}
G. Elmas, I. A. Litvin, and J. N\"otzel,
``Phase-Drift Limits and Adaptive Quadrature Readout in Programmable Photonic Processors,''
arXiv:2608.02249 (2026).

\bibitem{durt2010}
Durt, T., Englert, B. G., Bengtsson, I., \& Życzkowski, K. (2010). "On mutually unbiased bases." \textit{International journal of quantum information}, 8(04), 535-640.
\end{thebibliography}
\end{document}